\documentclass[10pt]{article}

\usepackage{amsmath}
\usepackage{amssymb}
\usepackage{lineno}

\usepackage{graphicx}

\usepackage{cite}

\usepackage{color} 

\date{}

\newcommand{\beq}{\begin{equation}}
\newcommand{\eeq}{\end{equation}}
\newcommand{\bbar}{\begin{eqnarray}}
\newcommand{\eear}{\end{eqnarray}}

\begin{document}

\begin{center}
{\LARGE
\textbf{Calculating effective gun policies}
}
\end{center}

\begin{center}
{\Large Dominik Wodarz$^{1,2,*}$ and  Natalia L. Komarova$^{2,1}$}
\end{center}

\noindent $^1$ Department of Ecology and Evolutionary Biology, 321 Steinhaus Hall, University of California Irvine, Irvine, CA 92697, USA. email: dwodarz@uci.edu

\noindent $^2$ Department of Mathematics, Rowland Hall, University of California Irvine, Irvine, CA 92697, USA. email: komarova@uci.edu

\bigskip 

\paragraph{Abstract.} Following recent shootings in the USA, a debate has erupted, one side favoring stricter gun control, the other promoting protection through more weapons. We provide a scientific foundation to inform this debate, based on mathematical, epidemiological models that quantify the dependence of firearm-related death rates of people on gun policies. We assume a shooter attacking a single individual or a crowd. Two strategies can minimize deaths in the model, depending on parameters: either a ban of private firearms possession, or a policy allowing the general population to carry guns. In particular, the outcome depends on the fraction of offenders that illegally possess a gun, on the degree of protection provided by gun ownership, and on the fraction of the population who take up their right to own a gun and carry it with them when attacked, parameters that can be estimated from statistical data. With the measured parameters, the model suggests that if the gun law is enforced at a level similar to that in the United Kingdom, gun-related deaths are minimized if private possession of firearms is banned. If such a policy is not practical or possible due to constitutional or cultural constraints, the model and parameter estimation indicate that a partial reduction in firearm availability can lead to a reduction in gun-induced death rates, even if they are not minimized. Most importantly,  our analysis identifies the crucial parameters that determine which policy reduces the death rates, providing guidance for future statistical studies that will be necessary for more refined quantitative predictions.


\section*{Introduction}

Gun violence has been an ongoing problem in the United States of
America, with an incidence of gun-related homicides that is
significantly higher than in most developed nations \cite{hemenway2000firearm}. While the pros and cons of gun control have been debated in the past, e.g. \cite{jacobs2002can}, this issue has recently gained new momentum.  On December 14,
2012, the country witnessed one of the worst school shootings in the
history of the US, where 20 children and 6 adults were killed in Sandy
Hook Elementary School in Newton, CT. This has sparked an intense
debate among politicians, interests groups, and media
personalities. On the one hand, this tragedy has resulted in a call
for tougher gun control laws. On the other hand, there is the
suggestion to arm the population in order to protect them against
offenders. This debate cannot be settled satisfactorily by verbal
arguments alone, since these can be simply considered opinions without
a solid scientific backing. What is under debate is essentially a
population dynamic problem: how do different gun control strategies
affect the rate at which people become killed by attackers,
and how can this rate be minimized? 

This question can be addressed with mathematical models that describe
the interaction between a criminal shooter and one or more people that
are the target of the shooter. The gun policy is defined as the
fraction of the population that can legally and readily obtain
firearms. On the one hand, the availability of firearms for a large
fraction of the population facilitates the acquisition of such weapons
by criminals, and this can increase the rate of attack on people. On the other hand, a relatively high prevalence of firearms in
the population can increase the chances of people to
successfully defend themselves against an attack, thus lowering the
death rate \cite{kleck1995armed, lott2010more}. The mathematical
models described in this study aim to analyze this tradeoff and to
suggest which type of gun policy minimizes firearm-related deaths under different assumptions. Calculations are performed for two
scenarios: the assault by a shooter of a single potentially armed
victim (what we call a one-on-one attack), or the assault of a crowd
of people that can be potentially armed (a one-against-many
attack). Note that the former scenario has been documented to be the
most prevalent cause of gun-related homicides \cite{cook1981effect,
  maxfield1989circumstances, fox2000homicide}. The latter scenario
corresponds to incidents such as movie theater or shopping mall
shootings and requires a more complicated model. Although such
one-against-many attacks are responsible for a small minority of
gun-related homicides, they are an important focus of public attention
and are widely discussed in the press.

According to our models, both sides of the gun control argument could
in principle work, depending on parameter values. Gun-related deaths can
be minimized either by the ban of private firearms possession, or by a
policy that allows the general public to obtain guns. The following crucial
parameters determine the optimal gun control policy: (i) the fraction
of offenders that cannot obtain a gun legally but possess one
illegally, (ii) the relative degree of protection against death
during an attack, conferred by gun ownership, and (iii) the fraction of people who take up their legal right to own a gun and carry it with them when attacked. These parameters can be estimated from published statistical data. In the context of the parameter estimates, the model suggests that if gun control laws
are enforced at a level similar to that in the United Kingdom, gun-related deaths can only be minimized by a ban of private firearm possession. If this policy is impractical for cultural or constitutional reasons, the parameterized model suggests that a partial reduction of firearm availability lowers the gun-induced death rate, even though it does not minimize it. Most importantly, the model identifies the crucial parameters that decide which policy reduces gun-induced deaths, providing a guide for what needs to be measured statistically in more detail.

\section*{Results}

To calculate the effect of different gun control policies on the
gun-induced death rate of people, we turn to the following
mathematical framework.  We consider the correlates of the total rate
(per year, per capita) at which people are killed as a result of
shootings.  We introduce the variable $g$ to describe the gun control
policy. This quantity denotes the fraction of the population owning a
gun. A ban of private firearm possession is described by $g=0$, while a "gun availability to all"
strategy is given by $g=1$. We assume that a certain small fraction of
the population is violent, and that an encounter with an armed
attacker may result in death. The number of offenders that own
firearms is a function of the gun control policy $g$ and is denoted by
$z(g)$. The probability of a person to die during an attack
is also a function of the gun control policy $g$ because this
determines whether the person and any other people also present at the
place of the attack are armed and can defend themselves. This
probability is denoted by $F(z)$. The overall risk of being killed by
a violent attacker as a result of shooting is thus proportional to
\beq
\label{main}
{\cal F}(g)=z(g)F(g).
\eeq
An important aspect of this model is the form of the dependency of
these two quantities on the gun control policy, $g$. The number of
armed attackers, $z(g)$, is a growing function of $g$, i.e. $z'>0$. Note,
however, that even if offenders are not allowed to legally obtain
firearms, there is a probability $h$ to obtain them illegally. Hence,
the value of $z$ is non-zero for $g=0$. One example of such a function
is given by the following linear dependence,
\beq
\label{z}
z(g)=g+h(1-g)
\eeq
with $0<h<1$. The probability
$F$ for a regular person to die in an attack (once he or she is at an
attack spot) is a decaying function of $g$, $F'<0$. One example again
is a linear function (see the following section). More generally,
again we could assume $F''>0$, see the one-against-many attack scenario below. 

We start by examining the case where the quantities $z$ and $F$ are
linear functions of the fraction of people that can legally obtain
firearms, $g$. This corresponds to the situation when a shooter attacks
a single individual in an isolated setting, i.e. no other people are
around to help defend against the attack. It could also correspond to
a classroom setting where a shooter attacks the entire class, but only
one person (the teacher) can be potentially armed for
protection. Subsequently, the more complex situation is examined where a
shooter attacks a group of people, each of which can be potentially
armed and contribute to defense. This would correspond to shootings in
movie theaters, malls, or other public places.

\subsection*{One-against-one attack}

Here we consider the situation where an attacker faces a single
individual who can be potentially armed. The fraction of people owning guns in the
population is defined by the legal possibility and ease at which guns
can be acquired ($g$), as well as the personal choice to acquire a
gun. Moreover, people who own a gun might not necessarily carry the firearm when attacked. Therefore, we will assume that the fraction of people armed with a gun when attacked is given by $cg$, where the parameter $c$ describes the fraction of people who take up their legal right of gun ownership and have the firearm in possession when attacked ($0\le c\le 1$). We will model the probability to be shot in an attack as
\beq
\label{Flin}
F(g)=\beta_1(1-cg)+\beta_2 cg,
\eeq
a linear function of $g$, where $\beta_1$ is the probability for an unarmed person to die in an
attack, and $\beta_2$ is the probability for an armed person to die in
an attack, with $\beta_1>\beta_2$. The number of attacks is
proportional to $z(g)$ given by equation (\ref{z}). The aim is to find the value of $g$ that minimizes the death rate, ${\cal F}$, given by equation (\ref{main}).

\paragraph{Optimization results.} The first important result is that the killing rate can only be
minimized for the extreme strategies $g=0$ and $g=1$, and that
intermediate strategies are always suboptimal (this is because ${\cal F}''<0$ for all $g$). In words, either a complete ban of private firearm possession, or a "gun availability to all" strategy minimizes gun-induced deaths. 

\begin{figure}
 \centering \includegraphics[scale=0.4]{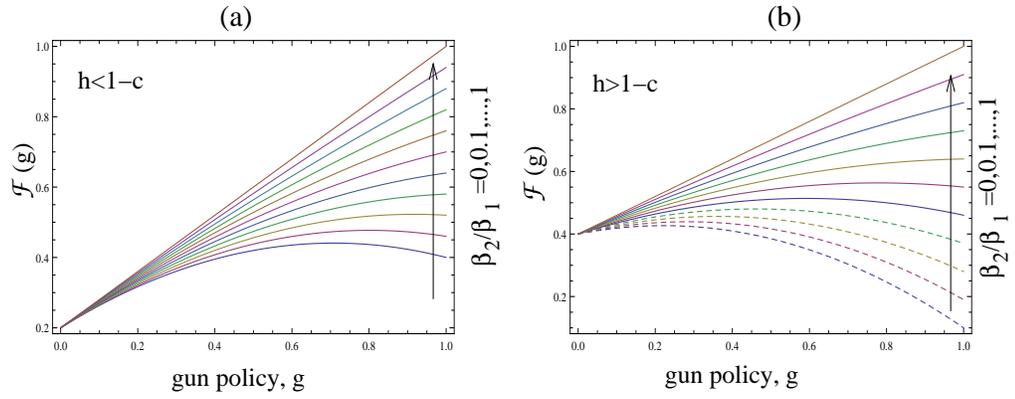}
   \vspace{\baselineskip}
   \caption{\footnotesize {\bf The rate of death caused by shooting in
       an one-against-one attack}, as a function of the gun control
     policy, $g$, where $g=0$ corresponds to a ban of private firearm
     possession, and $g=1$ to the ``gun availability to all''
     policy. (a) The fraction of people who possess the gun and have
     it with them when attacked is relatively low, $c=0.6<1-h$ with
     $h=0.2$. The different lines correspond to different values of
     $\beta_2/\beta_1$. For all values of $\beta_2/\beta_1$, the
     shooting death rate is minimal for $g=0$. (b) The fraction of
     people who possess the gun and have it with them when attacked is
     relatively high, $c=0.9>1-h$ with $h=0.4$.  As long as condition
     (\ref{in1}) holds, the shooting death rate is minimal for $g=0$
     (ban of private firearm possession, solid lines). If condition
     (\ref{in1}) is violated, then the shooting death rate is
     minimized for $g=1$ ("gun availability to all", dashed lines). }
\label{fig:one}
\end{figure}

Further, we can provide simple conditions on which of the two
extreme strategies minimizes death. Let us first assume that 
\beq
\label{hc}
h<1-c;
\eeq
this case is illustrated in figure \ref{fig:one}(a) and interpreted below.   Now, a ban of private firearm possession always minimizes gun-induced deaths. This can be seen in figure \ref{fig:one}(a), where we plot the shooting death rate, ${\cal F}$, as a function of the gun policy, $g$, for several values of the quantity $\beta_2/\beta_1$. For all of these functions, the minimum is achieved at $g=0$.  

Next, let us suppose that condition (\ref{hc}) is violated, that is, $h>1-c$, see figure \ref{fig:one}(b).  In this case, a ban of private firearm possession minimizes death if the following additional condition holds, 
\beq
\label{in1}
h<\frac{\beta_2}{\beta_1}c+1-c,
\eeq
Condition (\ref{in1})  defines a threshold value for $h$, the fraction of
offenders that cannot legally obtain a gun but possess one
illegally. If $h$ is smaller than the threshold value, then the policy
of choice is a ban of private firearm possession. The threshold value provided by inequality (\ref{in1}) depends on  the
degree to which gun ownership reduces the probability for the attacked
person to die, $\beta_2/\beta_1$ (the smaller the quantity
$\beta_2/\beta_1$, the higher the gun-mediated protection), and on the fraction $c$ of people who take up their right of legal gun ownership and carry the gun with them when attacked. The right hand side of inequality (\ref{in1}) decays with $c$. When $c=1$ (everybody who has a right to a gun, carries a gun), condition (\ref{in1}) takes a particularly simple form:
\beq
\label{in2}
h<\frac{\beta_2}{\beta_1}.
\eeq

The case where inequality (\ref{hc}) is violated is illustrated in figure \ref{fig:one}(b). Larger values of $\beta_2/\beta_1$ satisfy condition (\ref{in1}), see the solid curves in figure \ref{fig:one}(b). For those curves, $g=0$ (the total firearms ban) corresponds to the minimum of the shooting death rate. If condition (\ref{in1}) is not satisfied (the dashed curves in the figure), then the $g=1$ (``gun availability to all'')
policy is the optimum.  

Note that the key condition (\ref{hc}) relates two quantities, which in some sense are the
opposites of each other: The first quantity is $h$, the probability that a potential attacker who
cannot legally possess a gun will obtain it illegally and have it at
the time of the attack. This can be a measure of law enforcement, with lower values of $h$ corresponding to stricter law enforcement. The other quantity is $1-c$, the probability that a person who
can legally have a gun will not have it available when attacked. To make the ban of private firearm possession work (that is, to make sure that
it is indeed the optimal strategy), one would have to make an effort
to enforce the law and fight illegal firearm possession to decrease
$h$. To make the ``gun availability to all'' policy work, one would have to
increase $c$, for example by encouraging the general population to have
firearms available at all times. \\

\paragraph{Partial restriction of gun-ownership.} An important question is as follows. Let us suppose that the total gun 
ban is impossible due to e.g. constitutional or cultural constraints. Would a partial restriction of gun  ownership help 
reduce the firearm-related homicide rate? It follows that if
\beq
\label{2nd}
c<\frac{1-h}{(1-\beta_2/\beta_1)(2-h)},
\eeq
then any decrease in $g$ will reduce the gun-related homicide rate. If 
the value of $c$ is in the interval
$$\frac{1-h}{(1-\beta_2/\beta_1)(2-h)}<c<\frac{1-h}{1-\beta_2/\beta_1},$$
then the maximum death rate corresponds to an intermediate value of $g$ 
(while the minimum is at $g=0$, the total ban). This means that if the 
current state is $g=1$, then a partial reduction in $g$ may actually 
increase the gun-related homicide rate. The reduction must be 
significant, that is, $g$ has to be lowered below a threshold, in order 
to see a decrease in the gun-related death rate. Finally, if
$$c>\frac{1-h}{1-\beta_2/\beta_1},$$
which is the opposite of condition (\ref{in1}), then depending on 
$\beta_2/\beta_1$, the minimum of ${\cal F}$ may correspond to $g=1$.

\paragraph{A more general model of the victim population.} Equation (\ref{Flin}) 
defines the death probability of a person involved in an
attack. Comparing this equation with equation (\ref{z}) we can see
that in this model, we clearly separate the population of attackers
and the population of victims. The attackers carry a gun with
probability $P^{attacker}_{gun}=g+(1-g)h$, which assumes that the
attackers will obtain a gun if it is legally available, and have a
probability to also obtain it illegally. The victims carry a gun with
probability $P^{victim}_{gun}=cg$, which implies that they never
obtain a gun illegally, and even if it is legally available, they may
not have a firearm with them at the site of the attack. This could be
an appropriate model for homicide description in suburban and
low-crime areas. Below we will refer to this model as the ``suburban'' model. 

It is however possible that the population of victims is similar to
the population of the attackers in the context of gun ownership,
especially if we model the situation in different socio-economic
conditions, such as inner cities. The following model is more
appropriate for such situations: $P^{victim}_{gun}=cg+c_1(1-g)h_1$. It
states that a victim who is entitled to a legal weapon will have the
gun available at the time of the attack with probability $c$. Also,
victims who cannot possess a gun legally will acquire it illegally
with probability $h_1$ and have it with them with probability
$c_1$. This model reduces to model (\ref{Flin}) if $h_1=0$ (no illegal
gun possession among the victims). If on the other hand we set
$h_1=h$, then the population of victims is the same as the population
of attackers, apart from the fact that the attackers have a gun with
them with certainty (otherwise, there would be no attack), and the
victims may not be carrying a gun with them ($c,c_1<1$). We will refer
to this model as the ``inner-city'' model. In the follwoing text we
explore how our conclusions are modified under this more general
model.

The $g=0$ policy is the optimal as long as
\beq
\label{in3}
\frac{\beta_2}{\beta_1}>1-\frac{1-h}{c-c_1hh_1}.
\eeq
Note that if $h_1=0$, we have
$\frac{\beta_2}{\beta_1}>1-\frac{1-h}{c}$, which is the same as
condition (\ref{in1}). If $h_1=h$ and $c_1=c$, we have
$\frac{\beta_2}{\beta_1}>1-\frac{1}{c(1+h)}$, which is a weaker
condition than condition (\ref{in1}). In general, increasing $c_1$ and
$h_1$ makes condition (\ref{in3}) easier to fulfill. Therefore, we can
safely say that if condition (\ref{in1}) is fulfilled for the
``suburban'' model, then it will be fulfilled for the ``inner-city''
model. 

Furthermore, partial measures to reduce $g$ from $g=1$ will lead to a decrease in the death toll as long as 
$$\frac{\beta_2}{\beta_1}>1-\frac{1-h}{c(2-h)-c_1h_1}.$$
As before, with $h_1=0$ we recover condition (\ref{2nd}), and with an
increase in $h_1$ and $c_1$ lead to a weaker condition. Again, if
partial reduction of the gun ownership improves the death rate in the
``suburban'' model, it will also work in the ``inner-city'' model.

\subsection*{One-against-many attack}

Here, we consider a situation where a shooter attacks a crowd of people,
such as in a movie theater or mall shooting. The difference compared
to the previous scenario is that multiple people can potentially be
armed and contribute to stopping the attacker. We suppose that there
are $n$ people within the range of a gun shot of the attacker, and
$k$ of them are armed. We envisage the following discrete time Markov
process. At each time-step, the state of the system is characterized
by an ordered triplet of numbers, $(\alpha,i,j)$, where $\alpha\in
\{0,1\}$ tells us whether the attacker has been shot down ($\alpha=0$)
or is alive ($\alpha=1$), $0\le i\le k$ is the number of armed people
in the crowd, and $0\le j\le n-k$ is the number of unarmed people. The
initial state is $(1,k,n-k)$.

At each time-step, the attacker shoots at one person in the crowd
(with the probability to kill $0\le d\le 1$), and all the armed people
in the crowd try to shoot the attacker, each with the probability to kill
$0\le p\le 1$. The following transitions are possible from the state
$(1,i,j)$ with $0\le i\le k$, $0\le j\le n-k$ (below we use the convention that
expressions of type $i/(i+j)$ take the value $0$ for $i=j=0$):
\begin{itemize}
\item $(1,i-1,j)$: one armed person is shot, the attacker is not shot, with probability $d\frac{i}{i+j}(1-p)^{i-1}$;
\item $(1,i,j-1)$: an unarmed person is shot, the attacker is not shot, with probability  $d\frac{j}{i+j}(1-p)^{i}$;
\item $(0,i-1,j)$: one armed person is shot, the attacker is shot, with probability $d\frac{i}{i+j}[1-(1-p)^{i-1}]$;
\item $(0,i,j-1)$: an unarmed person is shot, the attacker is shot, with probability  $d\frac{j}{i+j}[1-(1-p)^{i}]$;
\item $(0,i,j)$: no potential victims are shot, the attacker is shot, with probability $(1-d)[1-(1-p)^{i}]$;
\item $(1,i,j)$: no one is shot, with probability  $(1-d)(1-p)^{i}$.
\end{itemize}
This model is considered in detail in the Analysis section. For $n>1$, $F(g)$ is a decaying function of $g$, with $F''> 0$ for all $n>1$. The following empirical model mimics the key properties of the overall risk of being shot in a one-against-many attack:
\beq
\label{mod}
{\cal F}=[g+(1-g)h]e^{-\beta cg+\gamma (cg)^2},
\eeq
where parameter $c$ is again the fraction of all the people who take up their legal right of gun ownership and carry the gun with them when attacked. The parameter $\beta$ measures the effectiveness of the protection
received from the guns, and parameter $0<\gamma<\beta/(2c)$ is used to better
describe the curvature of function $F(g)$ obtained from the exact
model. The empirical model (\ref{mod}) has the advantage of simplicity,
which allows for a straightforward analysis.

The optimal strategy that 
minimizes the gun-induced death rate of people again depends on the
degree of law enforcement (i.e. the probability for offenders to
obtain firearms illegally). More precisely, we have to evaluate the inequality 
\beq
\label{ineq2}
h<e^{-c\beta}.
\eeq
Again, the limiting case of this inequality corresponds to the case of $c=1$: 
\beq
\label{ineq}
h<e^{-\beta}.
\eeq
There are two cases:
\begin{itemize}
\item If inequality (\ref{ineq2}) holds (tight law reinforcement and/or gun protection ineffective), then the ``ban of private firearm possession'' policy ($g=0$) is optimal.
\item If $h>e^{-c\beta}$ (lax law reinforcement and/or gun protection highly effective), then depending on the value of $\gamma$ we may have different outcomes. Namely, if $\gamma<\frac{\beta}{2c}-\frac{1-h}{2c^2}$, then the $g=1$ (gun availability to all) policy is optimal. Otherwise, the optimal policy corresponds to an intermediate value of $g$:
$$g=\frac{\beta(1-h)-2c\gamma h+\sqrt{(\beta(1-h)-2c\gamma h)^2-8\gamma(1-h)(1-h-\beta ch)}}{4c\gamma(1-h)}<1.$$
\end{itemize}

\section*{Discussion}

We analyzed mathematical models in order to calculate the gun-induced
death rate of people depending on different gun control
strategies. The gun control strategies were expressed by a parameter
that describes the fraction of the population that can legally own
firearms. The strategies can range from a ban of private firearm possession to a
"gun availability to all" strategy. We first investigated a situation in which
one shooter is faced by only a single person that could potentially
own a gun and that could fight back against the shooter. This can
correspond to a one-on-one attack, such as a robbery, or a school
shooting where the only person in the classroom that could carry a gun
is the teacher. Subsequently, we examined a different scenario where a
shooter faces a crowd of people, all of which could potentially own a
gun and fight back against the attacker. This corresponds to shootings
in public places such as movie theaters and malls. The predictions of the model are similar for all scenarios. An
important parameter is the degree of law enforcement relative to the
amount of protection that gun ownership offers. If the
law is enforced strictly enough, a ban of private firearm possession minimizes the gun-induced death
rate of people. \\

The question arises how strict the law has to be enforced for the a ban of private firearm possession to minimize the gun-induced death of people. According to our results, this depends on the degree to which gun ownership protects potential victims during an attack and on the fraction of people who take up their legal right of gun ownership and carry the gun with them when attacked. These parameters in turn are likely to vary depending on the scenario of the attack and are discussed as follows. \\

\paragraph{One-against-one scenario.} The most prevalent use of guns is a one-against-one scenario and largely involves handguns \cite{maxfield1989circumstances, cook1981effect}. For this case, model predictions are relatively simple. Only one of the two extreme strategies can minimize gun-induced deaths, i.e. a ban of private firearm possession or a "gun availability to all" strategy. Intermediate gun policies lead to sub-optimal outcomes. Which strategy minimizes death depends on conditions that are easily interpreted.  Gun-induced deaths are always minimized by a firearm ban if $h<1-c$. That is, we have to compare $h$, the fraction of offenders that illegally own a gun, with $1-c$,  the fraction of the general population that could legally own a firearm but does not have it in possession when attacked.  If the condition above is not fulfilled, gun-induced deaths can be minimized by either strategy, depending on the fraction of offenders who illegally obtain firearms relative to the level of protection offered by gun-ownership during an attack. If $c=1$ (all people take up their right of gun ownership and carry it when attacked), the conditions is simplest, and a ban of private firearm possession minimizes gun-related deaths if $h<\beta_2/\beta_1$, where $\beta_2/\beta_1$ is inversely correlated with the degree of  protection offered by gun ownership to a victim during an attack, with $\beta_2/\beta_1=0$ meaning total protection, and $\beta_2/\beta_1=1$ corresponding to no protection associated with gun ownership. All these variables can be estimated from available statistical data, and the implications are discussed as follows. \\

In order to examine the fraction of offenders that cannot legally obtain a gun but own one illegally, $h$, we have to turn to a country with tough gun control laws. If a majority of people can legally own a gun, those that have to obtain one illegally is a negligible fraction. England and Wales have one of the strictest gun control laws since the 1997 Firearms Act, banning private possession of firearms almost entirely with the exception of some special circumstances \cite{firearmsact1997}. Estimating the fraction of potential offenders that illegally carry firearms is a difficult task. Most statistics quantify gun uses during the acts of offense, not among potential offenders. One study tried to fill this gap of knowledge by interviewing a pool of offenders that passed through prison \cite{bennett2004possession}. This was done in the context of the New English and Welsh Arrestee Drug Abuse Monitoring Programme (NEW-ADAM), covering a three year period between 1999-2002, and involving 3,135 interviewees. Among these offenders, 23\% indicated that they had illegally possessed a gun at some point in their life. However, only 8\% indicated illegal gun ownership within the last 12 months, which we consider a better measure of gun possession associated with committing crimes. More detailed questions revealed that only 21\% of people who owned a gun did so for the purpose of an offense. Similarly, among the 8\% of people who illegally owned a gun within the previous 12 months, only 23\% had taken the gun with them on an offense. Thus, as an estimate for the parameter $h$, we can say that 23\% of the 8\% constitutes people who illegally owned a gun which was also present during the offense, and hence $h=0.018$.  \\

The fraction of people who legally own a firearm and have it in possession when attacked, $c$, can be partially estimated. Statistical data are available about the fraction of people who personally own a gun in the United States, but no data are available that quantify the probability that these gun owners have the weapon with then when attacked. Approximately $30\%$ of all adult Americans own a gun \cite{miller2007household, gallup}. Because not all of them will have the firearm with them when attacked, we can say that $c<0.3$. In this scenario, a "gun availability to all" policy can minimize firearm-related deaths if $c>1-h$, i.e. if $0.3>0.982$. This condition is clearly violated, and so the model predicts that gun-related deaths are minimized by a ban of private firearm possession. It is possible that the fraction of offenders that illegally carry a firearm derived from the study by \cite{bennett2004possession} is an underestimate. This study analyzed gun ownership among a prison population. It is conceivable that among those offenders, only a certain proportion has sufficient violent intention, and that among those people the frequency of gun ownership is higher. According to our calculations, however, the "gun availability to all" policy can only minimize gun-induced deaths if more than $70\%$ of potential offenders illegally owned a firearm, i.e. if $h=0.7$ or higher, depending on how many of the legal gun owners are likely to carry the firearm when attacked. This is unlikely to be the case given the results obtained by \cite{bennett2004possession}, but needs to be investigated statistically in more detail.      \\

For the sake of the argument, let us consider the extreme scenario where all people who can legally own a gun do so and carry it with them at the time of an attack. This would require an effort by the government to persuade people to purchase firearms and carry them around at all times. As mentioned above, gun-related death is now minimized if $h<\beta_2/\beta_1$. The inverse relative protection that gun ownership provides during an attack ($\beta_2/\beta_1$) has also been statistically investigated \cite{branas2009investigating}. This is best done in a setting where a large fraction of the general population carries firearms, such as in the USA \cite{hemenway2000firearm}, and this study has been performed in Philadelphia. A total of 677 individuals assaulted with a gun were investigated and the study involved a variety of situations, including long range attacks where the victim did not have a chance to resist, and direct, short range attacks where the victim had a chance to resist. The study found that overall, gun ownership by potential victims did not protect against being fatally shot during an attack. In fact, individuals who carried a gun were more likely to be fatally shot than those who did not carry a gun. This also applies to situations where the armed victims had a chance to resist the attacker, and in this case, carrying a gun increased the chance to die in the attack about 5-fold. The authors provided several reasons for this. Possession of a gun might induce overconfidence in the victim's ability to fight off the attacker, resulting in a gun fight rather than a retreat. In addition, the element of surprise involved in an attack immediately puts the victim in a disadvantageous position, limiting their ability to gain the upper hand. If the victim produces a gun in this process rather than retreat, this could escalate the attacker's assault. These data would indicate that $\beta_2/\beta_1>1$, which in turn would again mean that a ban of private firearm possession is the only possible strategy that can minimize the gun-related death of people (the probability $h$ is by definition less than one). The results of this study have, however, been criticized on statistical grounds and it is currently unclear whether $\beta_2/\beta_1$ is indeed greater than one \cite{wintemute2010flaws, shah2010shah}. Results are also likely to depend on the geographical location. This study was conducted in a metropolitan area, and results might differ in smaller cities or more rural areas. However, the general notion that gun ownership does not lead to significant protection is also underlined by other studies that discussed the effectiveness of using guns as a defense against attacks \cite{kellermann1993gun, kleck1998risks, hemenway2000relative, hemenway2011risks}, especially in a home setting, although parameter estimates cannot be derived from these studies. In a literature review, no evidence was found that gun ownership in a home significantly reduces the chances of injury or death during an intrusion \cite{hemenway2011risks}. Even if this estimate of $\beta_2/\beta_1$ is somewhat uncertain, and even if its true value is less than one, the measured parameter $h=0.018$ means that firearm possession during an attack must reduce the chances of being fatally shot more than 50-fold for the "gun availability to all" policy to minimize the firearm-related deaths of people, i.e. the value of  $\beta_2/\beta_1$ must be less than 1/50. This is unlikely in the light of the above discussed studies and again points to a firearms ban for the general population to be the correct strategy to minimize gun-induced deaths.  \\

\begin{figure}
 \centering \includegraphics[scale=0.4]{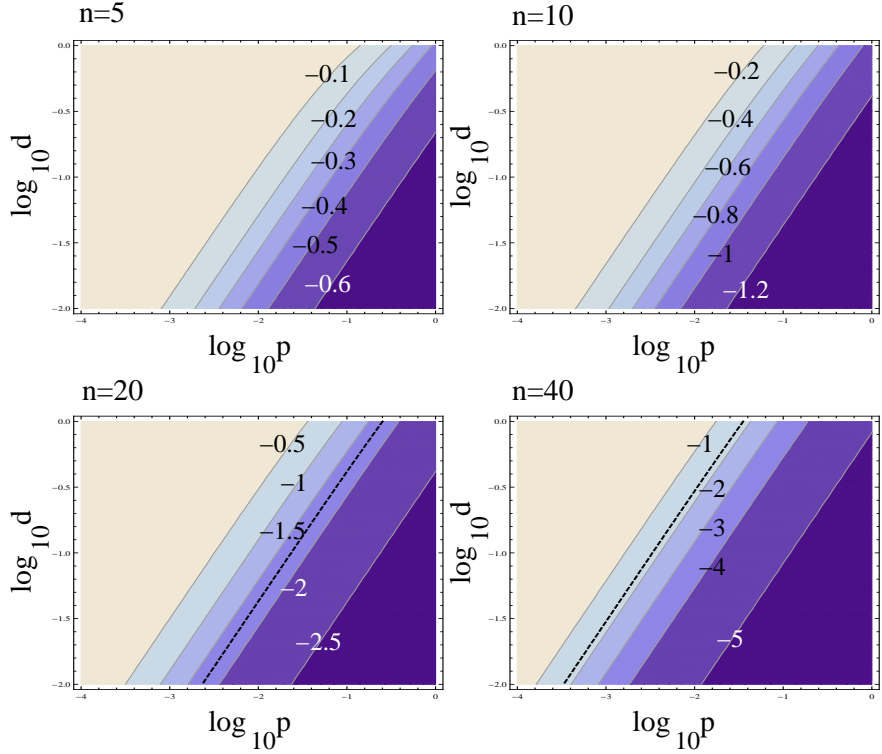}
   \vspace{\baselineskip}
   \caption{\footnotesize {\bf One-against-many attacks: when is a ban of private firearm possession the optimal policy?} Presented are the contour-plots of the threshold value $\log_{10}(e^{-c\beta})$ with $c=0.3$, as a function of $\log_{10}p$ and $\log_{10}d$ for four different values of $n$. Darker colors indicate smaller values, and the contour values are marked. For each pair of probabilities $p$ and $d$, the plots show the highest possible value of $\log_{10}h$ still compatible with the ban of private firearm possession being the optimal policy. The black dashed lines on the bottom two plots indicate the approximate location of the contour corresponding to $h=0.018$; above those lines the ban of private firearm possession is the optimal solution. These lines are drawn according to the following relationship between $d$ and $p$: $d=4p$ for $n=20$, and $d=30p$ for $n=40$. For $n=5$ and $n=10$, the inequality $h<e^{-c\beta}$ holds for any values of $p$ and $d$, and the ban of private firearm possession is the optimal solution in the whole parameter space. }
\label{fig:thr}
\end{figure}

It is important to note that while according to the measured parameters, a ban of private firearm possession minimizes the gun-induced death rate, the implementation of such a policy might be impractical in some countries like the USA, because  of constitutional and cultural constraints. It might only be practical to consider the option of partially restricting firearm access. In general, the model suggests that this might either decrease of increase the gun-induced death rate, depending whether condition  (\ref{2nd}) is fulfilled, which in turn depends on the measured parameter values. To interpret condition (\ref{2nd}), let us suppose that $h\ll 1$, as measured by \cite{bennett2004possession}. Then we have the condition
$$c<\frac{1}{2(1-\beta_2/\beta_1)}$$
that guarantees that any reduction in $g$ provides an improvement in safety. In the most extreme scenario where the gun-assisted protection is the highest ($\beta_2/\beta_1=0$), the threshold value of $c$ is given by 
$$c<\frac{1}{2}.$$
That is, as long as less than 50\% of people have the gun with them at
the time of an attack, a decrease in gun-ownership would decrease
gun-related homicides. With the estimate of $c\approx 0.3$ provided by statistical data, this condition holds. This means that even in the case
of very efficient weapons and high training of those who use them
($\beta_2/\beta_1=0$) a reduction in $g$ would be beneficial for the
society. In reality, the gun-related protection is not as high (which
corresponds to higher values of $\beta_2/\beta_1$), and the threshold
value of $c$ is most likely to be higher than $1/2$, which means that
most certainly a partial restriction on gun ownership would provide benefit
for reducing gun-related homicide rate.

\bigskip

\paragraph{One-against-many scenario.} Next, we discuss the one-against-many scenario. Here, two different gun control policies can again potentially minimize firearm-induced deaths of people: either a ban of private firearm possession, or arming the general population. However, in the latter case, not necessarily the entire population should carry firearms, but a certain fraction of the population, which is defined by model parameters. As in the one-against-one scenario, which policy minimizes gun-induced fatalities depends on the fraction of offenders that cannot legally obtain a gun but carry one illegally, the degree of gun-induced protection of victims during an attack, and the fraction of people who take up their right of gun ownership and carry the gun with them when attacked. In contrast to the one-against-one scenario, however, this dependence is more complicated here. In the empirical model described above, the degree of gun-mediated protection against an attack is given by the parameter $\beta$. This is a growing function of $n$ (the number of people involved in the attack) and $p$ (the probability for a victim to shoot and kill the attacker with one shot). Further, $\beta$  is a decaying function of $d$, the probability for the attacker to kill a victim with one shot. For a ban of private firearm possession to minimize gun-related deaths, the fraction of violent people that cannot obtain a gun legally but obtain one illegally must lie below the threshold given by condition (\ref{ineq}), i.e. $h<e^{-c\beta}$.  Let us again assume that $30\%$ of the general population owns a gun, and for simplicity that they all carry the firearm with them when attacked ($c=0.3$).  The
dependence of the function $e^{-c\beta}$ on the parameters $p$, $d$,
and $n$ is studied numerically in figure \ref{fig:thr}. Parameter $d$,
the probability for the attacker to kill a person with one shot,
varies between $10^{-2}$ and $1$. Parameter $p$, the probability for
an armed person to kill the attacker with one shot, varies between
$10^{-4}$ and $1$. The $\log_{10}$ of the right hand side of
inequality (\ref{ineq}) is represented by the shading, the lighter
colors corresponding to higher values. The dependence on parameters $p$ and $d$ is explored for different numbers of people in the crowd that is being attacked ($n=5, 10, 20, 40$). For each case, we ask what values of $p$ and $d$ fulfill the inequality $h<e^{-c\beta}$, assuming the estimated value $h=0.018$. In the parameter regions where this inequality holds, a ban of private firearm possession will minimize deaths, and outside those ranges, it is advisable that a fraction of the population is armed. For $n=40$ we find that a ban of private firearm possession requires $d>30p$. i.e. the attacker needs to be at least 30 more efficient at killing a victim than a single victim is at killing the attacker. For $n=20$ it requires $d>4p$. For $n=10$ and $n=5$, a firearms ban minimizes gun-related deaths for any value of $p$ and $d$.  Thus, for smaller crowds, condition (\ref{ineq}) is easily satisfied and a ban of private firearm possession would minimize deaths. For larger crowds, a ban of private firearm possession would only make sense if the attackers were significantly more efficient compared to the victims to deliver a fatal shot. The meaning of these numbers further depends on the weapon carried by the attacker. Strictly speaking, the model considered here was
designed for attackers and victims with similar weapons. The victims would typically possess hand guns. If the attacker also uses a hand gun, it can be questioned whether the attacker is 30 times more efficient at fatally shooting someone than a victim, even if the attacker is better trained and has more experience. Therefore, if most gun attacks in the country involved a one-against-many scenario with a crowd of about forty people or larger being assaulted by a non-automatic weapon, firearm-induced deaths would be minimized by a policy that allows a certain fraction of people to carry guns. For smaller crowds, a firearms ban is likely to minimize deaths.  In many situations, and certainly in the last string of mass shootings in the USA, automatic weapons were used to assault crowds, where hundreds of rounds per minute can be fired. The victims typically will not possess such powerful
weapons. Therefore, their ability to shoot is significantly lower than
that of the attacker.  We can interpret the results of the model for a situation where the
attacker fires a machine gun and the victims respond with
non-automatic weapons. In this case we must assume that the
probability of victims to fire and shoot the killer is significantly
(perhaps 2 orders of magnitude) lower than that of the attacker. In this case, a ban of private firearm possession minimizes gun-related deaths even if most  cases of gun violence involve the assault of relatively large crowds (such as $n=40$).

Now, for the sake of the argument assume that $c=1$, i.e. that everybody who can legally own a gun does so and has it in possession when attacked. The calculations yield the following results.  For $n=40$, a firearms ban requires that $d>125p$, i.e. the attacker needs to be at least 125 more efficient at killing a victim than a single victim is at killing the attacker. For $n=20$, $n=10$, and $n=5$, the conditions are $d>30p$, $d>6p$, and $d>0.6p$. While the numbers are now shifted in favor of a "gun availability to all" policy, the general conclusions hold. If most attacks in the country occur in a one-against-many setting involving large crowds where the attacker does not use an automatic weapon, firearm-related deaths are probably minimized by the "gun availability to all" policy. If the crowds under attack are relatively small or if the attacker uses an automatic weapon, gun-induced deaths are minimized by banning private possession of firearms. \\

Having discussed the one-against-many scenario in some detail, it has to be pointed out that while assaults on crowds generate the most dramatic outcomes (many people shot at once), the great majority of gun-related deaths occur in a one-against-one setting \cite{cook1981effect, maxfield1989circumstances, fox2000homicide}, which generates less press attention. Therefore, it is likely that the results from the simpler one-against-one scenario are the ones that should dictate policies when the aim is to minimize the overall gun-related homicides across the country. \\

\paragraph{Further complexities.} We have discussed the prediction of the model in the context of the parameter estimates using what we called the suburban model. That is, we separate attacker and victim populations. We have shown, however, that in the context of the inner city model, the condition for a ban of private firearm possession becomes easier to fulfill. This model does not separate the attacker and victim populations but instead describes a scenario where a large fraction of the population can have criminal tendencies, and where a person may either be an attacker or a victim depending on the situation, e.g. two armed people getting into a fight, drug related crimes, etc. Because the suburban model indicates that gun-related homicides are likely to be minimized by a ban of private firearm possession, the same conclusion would be reached with the inner city model. It has to be kept in mind though that parameter estimates could be different depending on the setting, although there is currently no information available about this in the literature. Related to this issue, it is clear that crime is not uniform with respect to spatial locations. There are areas with adverse socio-economic conditions which are characterized by high homicide rates, and there are areas of relative safety with very few gun crimes. While our model does not take space into account explicitly, it takes into account different scenarios (such as the suburbal model or the inner-city model). The optimization problem solved here does not explicitly depend on the spatial distribution of different crime conditions. Further questions about crime management can however be asked if one utilized a spatial extension of this model. \\

Another assumption that could be changed in the model is the dependency of $z(g)$ on the policy parameter $g$. In the model analyzed here, we assume that the fraction of offenders that are not entitled to have  a legal gun but  get it illegally, $h$, does not depend on $g$. Let us suppose that it does. In other words, as the prevalence of firearms in the general population increases, the fraction of criminals acquiring an illegal gun increases too. In this case, our general conclusions hold even stronger. As $g$ increases, the frequency of crime will increase even more, and the probability of death during an attack will remain the same as in the original model.\\

An issue that we have ignored in our discussion so far are possible deterring effects of gun ownership, i.e. the notion that non-homicide crimes, for example burglaries, could occur less often if those offenders are deterred by the presence of guns in households. Our analysis was concerned with minimizing gun-related homicides, and not crime in general, which is a different topic and should be the subject of future work. If a gun is present in households, and the burglar would consequently carry a gun during the offense, however, the number of gun-related deaths is likely to increase, even if perhaps the total number of burglaries might decrease. This applies especially if guns in the household are unlikely to protect against injury or death, as indicated in the literature \cite{hemenway2011risks}. \\

Finally, it is important to note that this paper only takes account of factors related to the gun control policy, and assumes a constant socio-economic background. Of course in the real world a reduction in gun-related (and other) homicides would require improvement of the living and work conditions and education of underprivileged populations. Here we do not consider these issues. It is important to emphasize that cultural and socio-economic differences exist between different countries, making it impossible to draw direct comparisons. A lower or higher rate of gun-related deaths is not only a function of gun control policies, but also of those other factors. 
Comparisons in the context of this model are only possible within the same cultural and socio-economic space. We single out the direct effects of gun-control policies and investigate those under fixed cultural and socio-economic circumstances.

\section*{Conclusions}

To conclude, this paper addresses the vigorous debate going on about the future of gun policies. One side argues that because guns are the reason for the deaths, stricter gun control policies should be introduced. The other side argues that guns also protect the life of  people, and that an increased prevalence of guns in the general population could lead to fewer deaths. Rather than relying on opinions, we investigated this debate scientifically, using mathematical, population dynamic models to calculate the firearm-induced death rate of people as a function of the gun control policy. The results show that in principle, both arguments can be correct, depending on the parameters. Based on parameters that could be estimated from previously published data, our model suggests that in the context of one-against-one shootings, a ban of private firearm possession minimizes gun-related deaths if the gun control law can be enforced at least as effectively as in England/Wales. However, if a private firearm ban is not practical or possible due to constitutional and cultural constraints, the model and parameter measurements suggest that a partial reduction in firearms availability could lower gun-induced death rates. In a one-against-many scenario, the situation is a little less clear. While in the context of assaults with automatic or semi-automatic weapons, gun ownership by the general population is unlikely to minimize firearm-induced deaths, it could be the right strategy when an attacker assaults a large crowd with a non-automatic weapon. The one-against-many scenario is, however, less important than the one-against-one scenario if the aim is to reduce the number of gun-related homicides in the country, because most gun-related homicides involve the attack of single individuals \cite{cook1981effect, maxfield1989circumstances, fox2000homicide}. The next step in this investigation is to perform further statistical studies to estimate the crucial parameters of the model that we have discussed here. These parameters are clearly identified by the model, which is perhaps the most important contribution of this analysis. Detailed statistical measurements would provide data that could further inform the design of gun control policies.

\section*{Analysis}

The absorbing states of the stochastic system are (i) all the states $(0,i,j)$, where the attacker has been shot, and (ii) $(1,0,0)$, where all the people have been shot. The state $(0,0,m)$ is unreachable from $(1,i,j)$. Let us denote by $h_{i,j\to l,m}$ the probability to be absorbed in state $(0,l,m)$ starting from state $(1,i,j)$. The goal is to calculate the function $F(g)$, which is proportional to the probability for an individual involved in an attack to die.

We have $h_{i,j\to l,m}=0$ as long as $l>i$ or $m>j$. Further, $h_{i,j\to 0,m}=0$ for all $(i,j,m)$. Let us denote by $\lambda_{i,j}$ the probability to be absorbed in state $(1,0,0)$ starting from state $(1,i,j)$. We must have
\beq
\label{check1}
\sum_{l=1}^i\sum_{m=0}^jh_{i,j\to l,m}+\lambda_{i,j}=1.
\eeq
The following equations can be written down for these variables, by using a one-step analysis:
\bbar
h_{i,j\to i,j}&=&(1-d)[1-(1-p)^i]+(1-d)(1-p)^ih_{i,j\to i,j},\nonumber \\
h_{i,j\to i-1,j}&=&\frac{di}{i+j}[1-(1-p)^{i-1}]+\frac{di}{i+j}(1-p)^{i-1}h_{i-1,j\to i-1,j}+(1-d)(1-p)^ih_{i,j\to i-1,j},\nonumber\\
h_{i,j\to i,j-1}&=&\frac{dj}{i+j}[1-(1-p)^i]+\frac{dj}{i+j}(1-p)^ih_{i,j-1\to i,j-1}+(1-d)(1-p)^ih_{i,j\to i,j-1},\nonumber\\
h_{i,j\to l,j}&=&\frac{di}{i+j}(1-p)^{i-1}h_{i-1,j\to l,j}+(1-d)(1-p)^ih_{i,j\to l,j},\quad l<i-1,\nonumber\\
h_{i,j\to i,m}&=&\frac{dj}{i+j}(1-p)^ih_{i,j-1\to i,m}+(1-d)(1-p)^ih_{i,j\to i,m},\quad m<j-1,\nonumber\\
h_{i,j\to l,m}&=&\frac{di}{i+j}(1-p)^{i-1}h_{i-1,j\to l,m}+\frac{dj}{i+j}(1-p)^ih_{i,j-1\to l,m}+(1-d)(1-p)^ih_{i,j\to l,m},\nonumber\\
&& \quad l<i,\quad m<j.\nonumber
\eear
The values for all $h_{ij\to lm}$ can be calculated recursively from
this system. For completeness, we also write down the equation for the
variable $\lambda_{i,j}$:
$$\lambda_{i,j}=\frac{di}{i+j}(1-p)^{i-1}\lambda_{i-1,j}+\frac{dj}{i+j}(1-p)^i\lambda_{i,j-1}+(1-d)(1-p)^i\lambda_{i,j}.$$
One can check that equation (\ref{check1}) holds.

\begin{figure}
 \centering \includegraphics[scale=0.22]{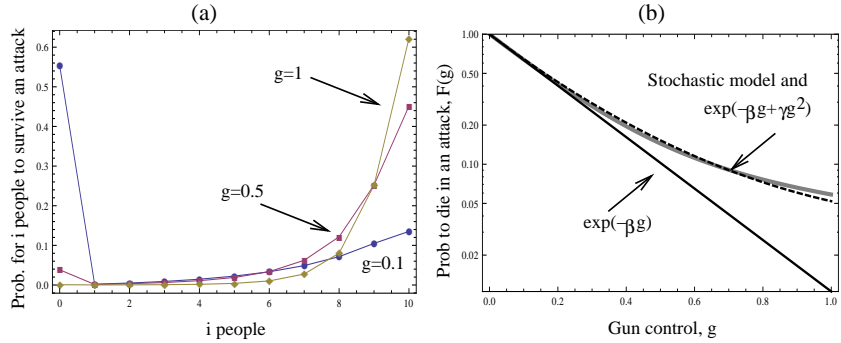}
   \vspace{\baselineskip}
   \caption{\footnotesize {\bf Results of the stochastic model of the one-against-many attack.} (a) A typical shape of the probability distribution of the number of people who survive an attack, plotted for three different values of $g$. (b) The probability to survive an attack, $F(g)$, as a function of the gun control (please note the logarithmic scale). The function $F(g)$ obtained in formula (\ref{exact}) is plotted by a thick gray line. The solid black line corresponds to approximation $e^{-\beta g}$ with $\beta$ given by formula (\ref{beta}). The dashed line corresponds to approximation (\ref{Fform}) with $\gamma=1.6$. Other parameters are $n=10$, $d=0.1$, $p=0.02$, and $c=1$. }
\label{fig:res}
\end{figure}

Given the probability to carry a gun $cg$, the probability to have $i$ individuals out of $n$ armed is given by 
$$\frac{n!}{i!(n-i)!}(cg)^i(1-cg)^{n-i},$$
and the probability to have $l$ armed and $m$ unarmed individuals still alive after an attack is given by
$$\sum_{i=1}^n \frac{n!}{i!(n-i)!}(cg)^i(1-cg)^{n-i} h_{i,n-i\to l,m}.$$
Therefore, the probability to have $k$ people out of $n$ to survive the attack is
$$P_k=\sum_{l=1}^k\sum_{i=1}^n \frac{n!}{i!(n-i)!}(cg)^i(1-cg)^{n-i} h_{i,n-i\to l,k-l}.$$
Examples of this probability distribution for three different values of $g$ are shown in figure \ref{fig:res}(a). The expected number of people that survive is given by
$$\sum_{k=1}^nP_kk.$$
Therefore, the function $F(g)$, proportional to the probability to be killed in an attack, is given by
\beq
\label{exact}
F(g)=1-\frac{1}{n}\sum_{k=1}^nP_kk.
\eeq

First of all, we can apply the one-against-many attack model  for the case $n=1$. As expected,  $F(g)$ for $n=1$ is a linear function of $g$, which can be written as equation (\ref{Flin}), with
parameters $\beta_1$ and $\beta_2$ giving rise to the threshold value of $h$,
$$\frac{\beta_2}{\beta_1}=\frac{d}{d(1-p)+p}.$$
We can see that the stochastic model informs our previous simple model by relating the quantity $\beta_2/\beta_1$ to the probability of the attacker to kill a victim with one shot, $d$, and the probability of a victim to shoot the attacker, $p$. As expected, the quantity $\beta_2/\beta_1$ grows with $d$ and decays with $p$. In other words, the gun-mediated protection decays with $d$ and it grows with $p$. 

The expression for $F$ for  $n>1$ is  complicated. A typical shape of this function is shown in figure \ref{fig:res}(b), the thick gray line. We can
calculate the approximation for this function for small values of
$g$, by setting
$$F(g)\approx 1-\beta cg,$$
where
\bbar
\beta&=&\frac{p}{n(d(1-p)+p)^n}\left(\sum_{j=1}^{n-1}j^2d^{n-j}(1-p)^{n-j-1}(d(1-p)+p)^{j-1}\right.\nonumber \\
&+&\left. n^2(1-d)(d(1-p)+p)^{n-1}\right)\nonumber \\
&=&\frac{1}{np^2}\left(2d^2+dp-2d^2p-2dnp+n^2p^2\right.\nonumber \\
\label{beta}
&-&\left.(2d(1-p)+p)d^{n+1}(1-p)^n(d(1-p)+p)^{-n}\right).
\eear

Instead of working with the particular model described above, let us design a simpler model, which would retain some of the properties of the stochastic model considered, but be easier to analyze. First we notice that $F'<0$ and $F''>0$. Consider the following approximation of this function which satisfies $F'<0$ and $F''>0$:
\beq
\label{appr}
F(g)\approx e^{-\beta cg}.
\eeq
Figure \ref{fig:res}(b) shows that while expression (\ref{appr}) is
a good approximation of the function $F(g)$ for small values of $g$,
it deviates from the function $F$ as $g$ approaches $1$, see the solid
black line. The function $F$ given by exact formula (\ref{exact})
has a higher curvature for larger values of $g$ (the thick gray line
in figure \ref{fig:res}(b)). To mimic this trend, we will set
\beq
\label{Fform}
F(g)=e^{-\beta cg+\gamma (cg)^2},
\eeq
where $\beta>0$ and $0<\gamma<\beta/(2c)$, such that
$F'<0$ for $0\le g\le 1$. This approximation is shown in figure
\ref{fig:res}(b), the dashed line. In equation (\ref{Fform}), 
$\beta$ is given by expression (\ref{beta}). 

If $\gamma=0$, we have ${\cal F}''<0$, and the optimal strategies are
the same as in the one-against-one attack: only the two extreme
strategies can minimize the gun-induced death rate of people, i.e. either
a "ban of private firearm possession" strategy ($g=1$) or the "gun availability to all" strategy
($g=1$). Conditions (\ref{ineq2}) (or (\ref{ineq}) if $c=1$) help
separate the two cases. If we assume the existence of a nonzero
correction $\gamma>0$ in the expression for $F(g)$, it follows that
inequality (\ref{ineq2}) still plays a key role in separating two
different cases, as described in the Results section.


\bibliography{guns}



\end{document}